\documentclass[runningheads]{llncs}

\usepackage[usenames,dvipsnames]{xcolor}
\usepackage[protrusion=true]{microtype}

\usepackage[english]{babel}
\usepackage{todonotes}
\usepackage[]{xcolor}
	
\usepackage{color, colortbl}
\definecolor{myteal}{rgb}{0.85, 0.98, 0.96}
\definecolor{myred}{rgb}{0.98, 0.85, 0.85}
\definecolor{mygrey}{rgb}{0.94, 0.94, 0.94}

\usepackage{multirow}

\usepackage{listings, xcolor}

\definecolor{verylightgray}{rgb}{.97,.97,.97}

\lstdefinelanguage{Solidity}{
	keywords=[1]{anonymous, assembly, assert, balance, break, call, callcode, case, catch, class, constant, continue, constructor, contract, debugger, default, delegatecall, delete, do, else, emit, event, experimental, export, external, false, finally, for, function, gas, if, implements, import, in, indexed, instanceof, interface, internal, is, length, library, log0, log1, log2, log3, log4, memory, modifier, new, payable, pragma, private, protected, public, pure, push, require, return, returns, revert, selfdestruct, send, solidity, storage, struct, suicide, super, switch, then, this, throw, transfer, true, try, typeof, using, value, view, while, with, addmod, ecrecover, keccak256, mulmod, ripemd160, sha256, sha3}, 
	keywordstyle=[1]\color{blue}\bfseries,
	keywords=[2]{address, bool, byte, bytes, bytes1, bytes2, bytes3, bytes4, bytes5, bytes6, bytes7, bytes8, bytes9, bytes10, bytes11, bytes12, bytes13, bytes14, bytes15, bytes16, bytes17, bytes18, bytes19, bytes20, bytes21, bytes22, bytes23, bytes24, bytes25, bytes26, bytes27, bytes28, bytes29, bytes30, bytes31, bytes32, enum, int, int8, int16, int24, int32, int40, int48, int56, int64, int72, int80, int88, int96, int104, int112, int120, int128, int136, int144, int152, int160, int168, int176, int184, int192, int200, int208, int216, int224, int232, int240, int248, int256, mapping, string, uint, uint8, uint16, uint24, uint32, uint40, uint48, uint56, uint64, uint72, uint80, uint88, uint96, uint104, uint112, uint120, uint128, uint136, uint144, uint152, uint160, uint168, uint176, uint184, uint192, uint200, uint208, uint216, uint224, uint232, uint240, uint248, uint256, var, void, ether, finney, szabo, wei, days, hours, minutes, seconds, weeks, years},	
	keywordstyle=[2]\color{teal}\bfseries,
	keywords=[3]{block, blockhash, coinbase, difficulty, gaslimit, number, timestamp, msg, data, gas, sender, sig, value, now, tx, gasprice, origin},	
	keywordstyle=[3]\color{violet}\bfseries,
	identifierstyle=\color{black},
	sensitive=false,
	comment=[l]{//},
	morecomment=[s]{/*}{*/},
	commentstyle=\color{gray}\ttfamily,
	stringstyle=\color{red}\ttfamily,
	morestring=[b]',
	morestring=[b]"
}

\usepackage{comment}

\usepackage{algorithm} 
\usepackage{algpseudocode} 
\usepackage{threeparttable,booktabs}

\usepackage{amssymb}
\usepackage{pifont}
\newcommand{\xmark}{\ding{55}}%

\usepackage{appendix}

\usepackage{amsmath}
\usepackage{graphicx}
\usepackage[colorlinks=true, allcolors=blue]{hyperref}
\usepackage{float}
\usepackage{booktabs}
\usepackage{cite}
\usepackage{siunitx}

\newcounter{mycounter}
\newcounter{itemcounter}[mycounter]

\renewcommand{\theitemcounter}{\thesection.\arabic{itemcounter}}

\newenvironment{myenv}[1]{%
  \begin{paragraph}{#1}%
    \setcounter{mycounter}{0}%
    \newcommand{\myitem}{\refstepcounter{itemcounter}\label{item\theitemcounter}\theitemcounter.~}%
}{%
  \end{paragraph}%
}

\begin{document}
\begin{sloppypar}
\title{Capturing Smart Contract Design\\with DCR Graphs}%
%
%
\author{Mojtaba Eshghie\inst{1}\orcidID{0000-0002-0069-0588}
    \and Wolfgang Ahrendt\inst{2}\orcidID{0000-0002-5671-2555}
    \and Cyrille Artho\inst{1}\orcidID{0000-0002-3656-1614}
    \and Thomas Troels Hildebrandt\inst{3}\orcidID{0000-0002-7435-5563}
    \and Gerardo Schneider\inst{4}\orcidID{0000-0003-0629-6853}
    }
\authorrunning{M. Eshghie et al.}
\titlerunning{Capturing Smart Contract Design with DCR Graphs}
\institute{KTH Royal Institute of Technology, Stockholm, Sweden\\ \email{eshghie@kth.se, artho@kth.se}
    \and Chalmers University of Technology, Gothenburg, Sweden\\ \email{ahrendt@chalmers.se}
    \and University of Copenhagen, Denmark\qquad\email{hilde@di.ku.dk}
    \and University of Gothenburg, Sweden\qquad\email{gerardo.schneider@gu.se}
    }
\maketitle
\definecolor{responseColor}{HTML}{1d8fff}
\definecolor{conditioncolor}{HTML}{ffa500}
\definecolor{milestoneColor}{HTML}{BA1FE5}
\definecolor{includeColor}{HTML}{29a719}
\definecolor{excludeColor}{HTML}{C10300}
\definecolor{noresponseColor}{HTML}{8c6026} 
\definecolor{valueColor}{HTML}{8c8c8c} 
\newcommand{\evalexp}[2]{\ensuremath{[[#1]]_{#2}}}
\newcommand{\DCR}{DCR\xspace}
\newcommand{\DCRR}{DCR$^*$\xspace}
\newcommand{\DCRL}{DCR$^\nu$\xspace}
\newcommand{\DCRB}{DCR$^!$\xspace}
\newcommand{\effectoff}[2]{\ensuremath{#1\cdot #2}}
\newcommand{\inputact}[2]{\ensuremath{?(#1, #2)}}
\newcommand{\outputact}[3]{\ensuremath{!(#1, #2, #3)}}
\newcommand{\outact}[2]{\ensuremath{!(#1, #2)}}
\newcommand{\intact}[2]{\ensuremath{*(#1, #2)}}
\newcommand{\ioactions}{\ensuremath{\mathsf{IO_{A,D}}}}
\newcommand{\inoutactions}{\ensuremath{\mathsf{IO_{A}}}}
\newcommand{\interact}[3]{\ensuremath{(#1, #2 \rightarrow #3)}}
\newcommand{\actt}[3]{\ensuremath{#1(#2,#3)}}
\newcommand{\interactdata}[4]{\ensuremath{(#1, #2 \stackrel{#3}{\longrightarrow} #4)}}
\newcommand{\rref}{\sqsubseteq}
\renewcommand\t{\ensuremath{\mathsf{t}}}
\newcommand\f{\ensuremath{\mathsf{f}}}
\newcommand{\VAL}{\ensuremath{V}} 
\newcommand{\natinf}{\ensuremath{\infty}}

\newcommand{\lab}{\ensuremath{l}}
\newcommand{\valuerel}{{\color{valueColor} \ensuremath{\mathrel{\rightarrow\!\!=}}}}
\newcommand{\conditionrel}{{\color{conditioncolor} \ensuremath{\mathrel{\rightarrow\!\!\bullet}}}}
\newcommand{\responserel}{{\color{responseColor}\ensuremath{\mathrel{\bullet\!\!\rightarrow}}}}
\newcommand{\milestonerel}{{\color{milestoneColor} \ensuremath{\mathrel{\rightarrow\!\!\diamond}}}}
\newcommand{\includerel}{{\color{includeColor} \ensuremath{\mathrel{\rightarrow\!\!\textsf{+}}}}}
\newcommand{\excluderel}{{\color{excludeColor} \ensuremath{\mathrel{\rightarrow\!\!\textsf{\%}}}}}

\newcommand{\noresponserel}{{\color{noresponseColor}\ensuremath{\mathrel{\bullet\!\!\!\rightarrow\!\!\!\times}}}}

\newcommand{\gconditionrel}[2]{{\color{conditioncolor}\ensuremath{\stackrel{[#1]}{\mathrel{\rightarrow\!\!\bullet}}_{d}}}}
\newcommand{\gresponserel}[2]{{\color{responseColor}\ensuremath{\stackrel{[#1]}{\mathrel{\bullet\!\!\rightarrow}}_{#2}}}}
\newcommand{\gmilestonerel}[1]{{\color{milestoneColor}\ensuremath{\stackrel{[#1]}{\mathrel{\rightarrow\!\!\diamond}}}}}
\newcommand{\gincluderel}[1]{{\color{includeColor}\ensuremath{\stackrel{[#1]}{\mathrel{\rightarrow\!\!+}}}}}
\newcommand{\gexcluderel}[1]{{\color{excludeColor}\ensuremath{\stackrel{[#1]}{\mathrel{\rightarrow\!\!\%}}}}}
\newcommand{\gnoresponserel}[1]{{\color{noresponseColor}\ensuremath{\stackrel{[#1]}{\mathrel{\bullet\!\!\!\rightarrow\!\!\!\times }}}}}

\newcommand{\responses}{\ensuremath{\mathsf{Re}}}
\newcommand{\executed}{\ensuremath{\mathsf{Ex}}}
\newcommand{\included}{\ensuremath{\mathsf{In}}}

\newcommand{\markingset}{\ensuremath{\mathcal{M}}}
\newcommand{\graphs}{\ensuremath{\mathcal{G}}}
\newcommand{\pgraphs}{\ensuremath{{\cal{P}}}}

\newcommand{\genrel}{\ensuremath{\mathrel{\rightarrow}}}
\newcommand{\genrelto}[1]{\genrel\!#1}
\newcommand{\genrelfrom}[1]{#1\!\genrel}
\newcommand{\enable}[2]{\ensuremath{\mathsf{enabled}(#1,#2)}} 
\newcommand{\power}[1]{\ensuremath{{\cal P}(#1)}}
\newcommand{\maxc}{\ensuremath{maxc_G}}
\newcommand{\minr}{\ensuremath{minr_G}}
\newcommand{\miner}{\ensuremath{minRe_G}}
\def\L{\mathsf{L}}

\def\lEx{\mathsf{L_{Ex}}}
\def\lRe{\mathsf{L_{Re}}}
\def\lIn{\mathsf{L_{In}}}
\def\dom{\mathsf{dom}}

\newcommand{\spg}{\ensuremath{sp}} 
\newcommand{\ESG}{\ensuremath{E}}

\newcommand\Ex{\mathsf{Ex}}
\renewcommand\Re{\mathsf{Re}}
\newcommand\In{\mathsf{In}}

\newcommand\Va{\mathsf{Va}}

\newcommand{\commitevent}{\ensuremath{\mathsf{commit}}}
\newcommand{\revealevent}{\ensuremath{\mathsf{reveal}}}
\newcommand{\transactionevent}{\ensuremath{\mathsf{revealtransaction}}}

\newcommand{\draftcomment}[3]{{\color{#1}[#3] #2} 
  \PackageWarning{WARNING: Draft comments visible}{#2: #3}}
\newcommand{\gs}[1]{\draftcomment{red}{GS}{#1} }
\newcommand{\wa}[1]{\draftcomment{blue}{WA}{#1} }

\begin{abstract}
Smart contracts manage blockchain assets and embody business processes.
However, mainstream smart contract programming languages such as Solidity lack explicit notions of roles, action dependencies, and time.
Instead, these concepts are implemented in program code. This makes it very hard to design and analyze smart contracts.

We argue that DCR graphs are a suitable formalization tool for smart contracts because they explicitly and visually capture the mentioned features. 
We utilize this expressiveness to show that many common high-level design patterns representing the underlying business processes in smart-contract applications can be naturally modeled this way.
Applying these patterns shows that DCR graphs facilitate the development and analysis of correct and reliable smart contracts by providing a clear and easy-to-understand specification.

\keywords{Smart Contract Modelling\and DCR Graphs\and Design Patterns}
\end{abstract}

\section{Introduction}\label{sec:intro}

A \emph{smart contract} is implemented as immutable code executed on a blockchain and
may be seen as 
a special business process specifying a contractual agreement on actions to be carried out by different roles.
While smart contracts offer advantages such as uncompromised (automated) execution even without a trusted party,
they can also be complex and difficult to design and understand. This is even more problematic as they cannot be changed once deployed.



In a normal business process environment, different roles collaborate
to achieve a common business goal. In contrast, different roles in a smart contract typically have adversarial interests. Therefore, smart contracts introduce new types of patterns of behavior,
which have so far only been informally described~\cite{EmpiricalAnalysisSmart2017,DesignPatterns2018,securityPatterns,SolidityPatternsWebsite}. To provide an unambiguous understanding of the patterns that can also provide the basis for formal specifications, we set out to extend the study and formalization of process patterns to include these smart contract patterns.

Solutions to adversarial-interest problems often use time- or data-related constraints between actions cutting across the process and the more standard use of roles and sequential action dependencies.  We find that a declarative notation involving data and time is appropriate for formalizing the new smart contract process patterns. Moreover, smart contract languages exhibit a transactional behavior of actions, where an action may be attempted but aborted if the required constraints for executing it are not fulfilled. This suggests that individual actions have a life cycle, like sub-processes.

For these reasons, we use DCR graphs~\cite{dcrdata,dcrsub}, which are by now a well-established declarative business process notation that has been extended with data~\cite{dcrdata}, time~\cite{dcrdata}, and sub-processes~\cite{dcrsub}.
%
DCR graphs visually capture important properties such as the partial ordering of events, roles of contract users, and temporal function attributes. Using DCR graphs, it is possible to represent a smart contract with a clear and concise model that is more expressive and comprehensive than other types of models. 
As the design patterns we model concern the \emph{high-level} behavior of a smart
contract under analysis, we elide technical details of the patterns' implementation and execution. Therefore, we use the term ``high-level'' design pattern for the patterns that DCR graphs capture well, as they represent the underlying business process of the contract.
Further, DCR models are useful for analysis. We show that using DCR graphs facilitates the development of correct and reliable smart contracts by providing a clear and easy-to-understand specification. 
More concretely, our contributions are:
\begin{enumerate}
    \item We systematically identify and distinguish high-level design patterns from low-level (implementation-specific) patterns in smart contracts  (Table \ref{table:patterntable}), and demonstrate how we model them with DCR graphs by going through four of the most complex ones (\S \ref{time-incent-dcr}, \S \ref{rate-limiting-dcr}, \S \ref{commit-and-reveal-dcr}, and \S \ref{circuit-breaker-dcr}). The DCR models of the rest of the 19 patterns may be found in the accompanying repository \cite{SolidityDesignPatternsModeling}.  
    
    \item We demonstrate how one can capture the design of a complete contract, not just a design pattern, with the help of DCR graphs (casino example in Section \ref{sec:casino}).
    The modeled contract has three of the design pattern models from this paper incorporated, which helps to demonstrate the combinability of pattern models to shape the final design of the contract. 

    \item As a result of a thorough analysis of real-world contracts, including popular contract libraries, we identify (and model) two new design patterns: \emph{time incentivization} (\S \ref{time-incent-dcr}) and \emph{escapability} (\S \ref{escapability-dcr}). Both of these patterns are extensively used by the Solidity developer community but are not yet introduced as design patterns in research literature \cite{securityPatterns, DesignPatterns2018, designPatternsGas, SolidityPatternsWebsite}. 
\end{enumerate}
Our application of these formalized design patterns in Section \ref{sec:casino} shows that using DCR graphs can facilitate the development of correct and reliable smart contracts by providing a clear and easy-to-understand specification. Moreover, DCR specifications can provide a basis for automated (dynamic or static) analysis of smart contracts, which we exemplified by preliminary runtime verification infrastructure and experiments in our tool paper \cite{clawk}. 

Our usage of DCR graphs to model smart contracts and our focus on high-level rather than low-level properties allows us to capture the key semantics of the contract succinctly. We can verify properties (and likewise lack of vulnerabilities pertaining to these properties) related to roles and access control \cite{SWC105, SWC106}, partial ordering of actions (function calls and transaction execution) \cite{SWC114}, as well as time-based vulnerabilities \cite{TimestampDependenceEthereum, SWC116}. Furthermore, not being concerned with low-level patterns and properties lets our approach remain cross-platform and not tied to the features and limitations of a certain smart contract execution environment. We believe that these patterns provide a systematic classification of best practices for smart contracts in a similar way that software design patterns shaped the design of traditional software and established a nomenclature for it~\cite{gamma1995design}, while capturing aspects that are unique to smart contracts.

This paper is organized as follows: Section~\ref{sec:background} introduces smart contracts and DCR graphs. Section~\ref{sec:dcr-patterns} gives an overview of 19 smart contract design patterns, which we formalize as DCR graphs.
Section~\ref{sec:casino} shows a case study on a casino smart contract. Section~\ref{sec:related-work} covers related work, and Section~\ref{sec:conclusion} concludes.

\section{Background}\label{sec:background}
\subsection{Smart Contracts: Ethereum and Solidity}\label{sec:preli:sol}

Ethereum~\cite{wood2014ethereum}, with its built-in cryptocurrency Ether, is
still the leading blockchain framework supporting smart contracts.
In Ethereum, not only the users but also the contracts can receive, own, and send Ether. 
Ethereum miners look for transaction requests on the network, which
contain the contract's address to be called, the call data, and the amount of Ether to be sent. Miners are paid for their efforts in (Ether priced) \emph{gas,} to be paid by the initiator of the transaction.
\begin{figure}[t]
\begin{lstlisting}[basicstyle=\fontsize{8}{10}\ttfamily,numberstyle=\tiny,language=Solidity,escapechar=|]
contract Casino {
  address public operator, player; bytes32 public hashedNumber;
  enum State { IDLE, GAME_AVAILABLE, BET_PLACED };
  State private state;  uint bet;  Coin guess;
  function addToPot() public payable byOp {...|\label{func:addToPot:add}|}
  function removeFromPot(uint amt) public|\label{func:removefrompot}| byOp, noActiveBet|\label{func:removeFromPot:modifier}| {...|\label{func:removeFromPot:removeAmount}|}
  function createGame(bytes32 hash) public|\label{func:createGame}| byOp, inState(IDLE) |\label{func:createGame:modifier}|{
    hashedNumber = hash|\label{func:createGame:hashednumberAssign}|
    state = GAME_AVAILABLE;|\label{func:createGame:gameavailable}|}
  function bet(Coin _guess) public payable|\label{func:placeBet}| inState(GAME_AVAILABLE)|\label{func:placeBet:modifier}| {
    require (msg.sender != operator);|\label{func:placeBet:requireNotOper}|
    require (msg.value > 0 && msg.value <= pot);|\label{func:placeBet:requireValue}|
    player = msg.sender; bet = msg.value;
    guess = _guess; state = BET_PLACED;|\label{func:placeBet:stateBP}|}
  function decideBet(uint secret) public|\label{func:decideBet}| byOp, inState(BET_PLACED) |\label{func:decideBet:modifier}|{
    require (hashedNumber == keccak256(secret));
    Coin secret = (secret% 2 == 0)? HEADS : TAILS;|\label{func:decideBet:secretAssign}|
    if (secret==wager.guess) {playerWins();} else {operatorWins();}
    state = IDLE;|\label{func:decideBet:goToInit}|}
}
\end{lstlisting}
\caption{Solidity-code for casino (some details are omitted)}
\label{fig:solcode1}
\end{figure}


A transaction is not always executed successfully. It can be reverted due to running out of gas, sending of unbacked funds, or failing runtime assertions. If a miner attempts to execute a 
transaction, a revert statement within the transaction's execution can \emph{undo the entire transaction.}
All the effects so far are undone (except for the paid gas), as if the original call had never happened.

The most popular programming language for Ethereum smart contracts is
Solidity~\cite{solidity-web}. 
Solidity follows largely an object-oriented paradigm, with
fields and methods, called `state variables' and `functions',
respectively. Each external user and each contract instance has a unique
\lstinline|address|. Each \lstinline|address| owns Ether (possibly 0),
can receive Ether, and send Ether to other addresses. For instance,
\lstinline[mathescape=true]|$a$.transfer($v$)| transfers an amount $v$
from the caller to $a$.

The current caller, and the amount sent with the call, are always available via \lstinline|msg.sender| and  \lstinline|msg.value|, respectively. Only \lstinline|payable| functions accept payments.
Fields marked \lstinline|public| are read-public, not write-public. Solidity also offers some cryptographic primitives, like \lstinline|keccak256| for computing a crypto-hash. \lstinline[mathescape=true]{require($b$)} checks the Boolean expression $b$, and reverts the transaction if $b$ is false. 

Solidity further features programmable \emph{modifiers}. The contract in Fig.~\ref{fig:solcode1} uses the modifiers \lstinline|byOp|, \lstinline|inState(s)|, and \mbox{\lstinline|noActiveBet|,} whose implementation is omitted for brevity. These three modifiers expand to \lstinline[mathescape=true]{require($b$)}, where $b$ is \lstinline|msg.sender == operator|, \lstinline|state == s|, and \lstinline|state != BET_PLACED|, respectively.

\subsection{Dynamic Condition Response Graphs}
\begin{figure}[t]
    \centering \includegraphics[scale=0.57]{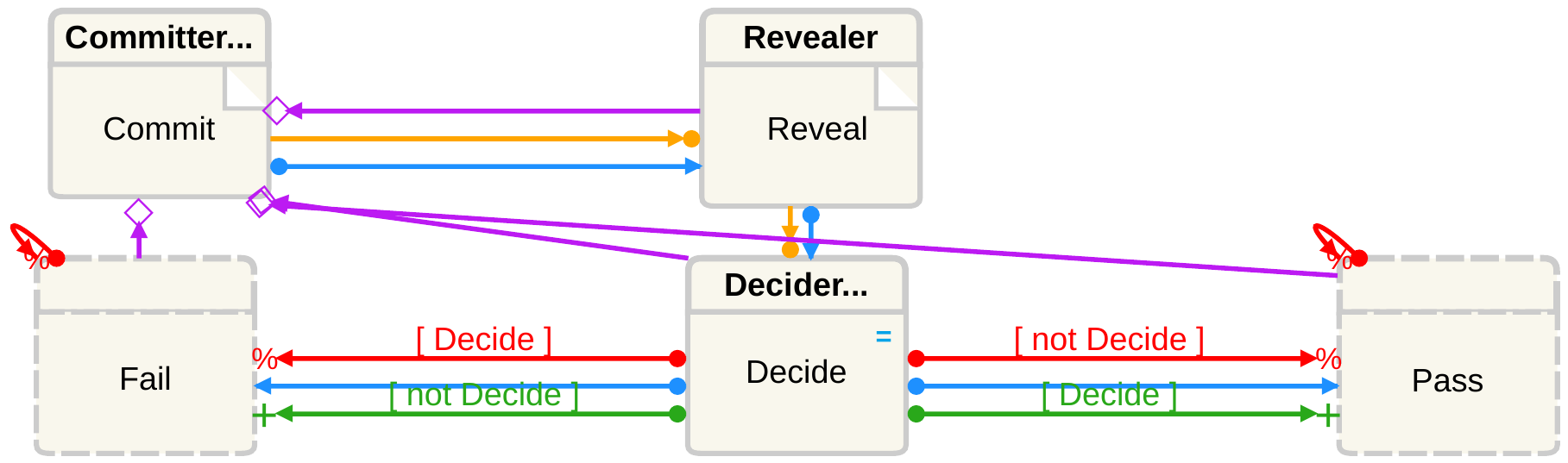}
    \caption{Commit and reveal design pattern}  
    \label{fig:commit-and-reveal} 
\end{figure}

A dynamic condition response (DCR) graph defines a dynamic process declaratively as a graph, defined formally in Def.~\ref{def:timedDCR} below and exemplified in Fig.~\ref{fig:commit-and-reveal}. 
DCR graphs offer an alternative to state machines; instead of using transitions to represent events, DCR graphs represent events as nodes (boxes). Events in a DCR graph may be restricted to certain roles. Events can be enabled or disabled by other events, which is represented by different types of arrows.

The nodes of the graph constitute a set $E$ of events labeled with roles and an action, visualized in Fig.~\ref{fig:commit-and-reveal} as boxes with the action label in the middle and the role label in the top bar. Nodes can be either input actions (denoted by a flipped paper corner in the top right of the box containing the action label; in this example, actions \emph{commit} and \emph{reveal}), computation actions (denoted by an $=$-sign in the top right of the box containing the action label; in this example, the \emph{decide} action) or simple actions (in this example, the \emph{fail} and \emph{commit} actions).
Input actions receive a value from the environment when the action is executed, which is associated with the event. Computation actions execute a computation expression (that may refer to the current value assigned to itself or other events) when the action is executed, which is then associated with the event. In this example, the computation assigned to the \emph{decide} action is the Boolean expression \emph{commit = hash(reveal)} (not shown graphically in Fig.~\ref{fig:commit-and-reveal}), which refers to the values of the \emph{commit} and the \emph{reveal} actions.

The directed edges between nodes define rules for the execution of events. The rules can be constraints or effects. An example of a constraint is the condition rule, visualized in Fig.~\ref{fig:commit-and-reveal} as an orange arrow $\conditionrel$ with a bullet at the target. It states that the event at the source of the edge (in this example, the commit action) must have been executed at least once (or be excluded) for the event at the target (in this example, the \emph{reveal} action) to be executable. 

Examples of effects are the \emph{exclude}, \emph{include} and \emph{response} rules, visualized in Fig.~\ref{fig:commit-and-reveal} as respectively a red arrow $\excluderel$ with a $\%$-sign at the target, a green arrow $\includerel$ with a $+$-sign at the target, and a blue arrow $\responserel$ with a dot at the source. The exclude (include) rule states that when the event at the source (in this case, the \emph{decide} action) is executed, the events at the target (in this case, the \emph{fail} and \emph{pass} actions) are excluded (included). 
Excluded events cannot be executed and are also ignored when determining constraints. The possibility for an event to be excluded makes it easy to express \emph{defeasible rules}~\cite{defeasiblelogicNute}. For instance, in Fig.~\ref{fig:time-incent-dcr}, the bank can give a fine a month after a loan, except if the client, in the meantime, pays the loan, in which case the event of the fine action is excluded.

In DCR graphs with data, rules may be guarded by Boolean expressions, determining whether a rule is to be considered in the current state of the graph. In this example, the guard \emph{decide} of the exclude relation $\excluderel$ from \emph{decide} to \emph{fail} means that \emph{fail} is excluded if and only if the value of \emph{decide} is true, which is the case if the committed value provided when \emph{commit} is executed is equal to the hash of the value provided when \emph{reveal} is executed. The response rule $\responserel$ denotes that if the event at the source (e.\,g., the \emph{commit} action in Fig.~\ref{fig:commit-and-reveal}) is executed, then the event at the target (e.\,g., the \emph{reveal} action in Fig.~\ref{fig:commit-and-reveal}) must be executed or excluded in the future.

The execution state of a DCR graph is given by a marking, which assigns state information to each event. In the original version of DCR graphs~\cite{Hildebrandt2011}, the marking of the graph assigned three Booleans to each event, denoting respectively if the event had been executed, if it is required to be executed (or excluded) in the future and if it is currently excluded. 
In this paper, we use an extended version of DCR graphs, allowing both data, time and nested sub-processes, which is supported by the online design tool.\footnote{Available for free for academic use at \url{dcrsolutions.net}} This version of DCR graphs also adds two new effect rules: A \emph{value relation} $\valuerel$, denoted by a grey arrow with an $=$-sign at the target, with the effect of updating the value of the target event when the source event is executed, and a \emph{cancel relation} $\noresponserel$, denoted by a brown arrow with a $\times$-sign at the target, with the effect of removing a possible pending execution requirement (e.\,g., due to a previous activation of a response rule) of the target event when the source event is executed.

For a DCR graph with data, the marking assigns the current data value (if any) associated with each event, as exemplified above. 
For a DCR graph with time, the marking additionally assigns time information to events, concretely, how long ago an event was executed (if it has been executed) and a deadline for when it is required to be executed (if it is required to be executed in the future). 

In Def.~\ref{def:timedDCR}, we give the formal definition of timed DCR graphs with sub-processes and data. We combine timed DCR graphs with sub-processes~\cite{dcrsub} and timed DCR graphs with data~\cite{dcrdata} and add a new type of edge denoting a value effect, making it possible for one event to update the data of another event. 

We assume a set of computation expressions $\mathsf{Exp_E}$, with $\mathsf{BExp_E}\subseteq \mathsf{Exp_E}$ being a subset of Boolean expressions.
 For every event $e\in E$, we assume an expression $e\in \mathsf{Exp_E}$ that denotes the current value of the event (as recorded in the marking). We also assume a discrete-time model (i.\,e., time is represented as time steps given as natural numbers) and let $\omega$ denote the natural numbers (including 0) and $\infty=\omega \cup \{\omega\}$, i.\,e, the natural numbers and $\omega$ (infinity).\footnote{The ISO 8601 standard (\url{www.iso.org/iso-8601-date-and-time-format.html}) is used in the design tool, allowing the use of years, months, days, and seconds.} Infinity is used to represent a non-fixed deadline of a required event, i.\,e., that an event must eventually be executed as known from classical liveness properties. This is the default deadline of a response relation if the deadline is not given, as it is the case for the two response relations in 
 Fig.~\ref{fig:commit-and-reveal}.

\begin{definition}
\label{def:timedDCR}
    A \emph{timed DCR graph with sub-processes, data, and roles} $G$ is given by a tuple 
    $(E, \spg, D, M, \conditionrel, $ $ \responserel, \noresponserel, \milestonerel, \includerel, \excluderel, \valuerel, L,l)$
    where 
    \begin{enumerate}
    \item $E$ is a finite set of \emph{events}, 
   \item $\spg\in \ESG\rightharpoondown \ESG$ is an acyclic \emph{sub-process} function,  i.\,e., for all $k>1$
$\spg^k(e)\not = \spg(e)$, if $\spg(e)$ is defined.
 \item $D:E\rightarrow \mathsf{Exp_E}\uplus\{?\}$ defines an event as either a \emph{computation event} with expression $d\in \mathsf{Exp_E}$ or an \emph{input event} $?$,
    \item  $M = (\Ex,\Re,\In,\Va) \in  \bigl(
        {(E\rightharpoondown \omega)
        \mathrel\times
        {(E\rightharpoondown\infty)}
        \times
        \mathcal{P}(E)
        \times (E\rightharpoondown \VAL)}
        \bigr)$ 
    is the \emph{timed marking with data}, 
     \item $\conditionrel \subseteq E \times \omega \times \mathsf{BExp_E} \times E$, is the \emph{guarded timed condition relation}, 
     \item $\responserel \subseteq E \times \infty \times \mathsf{BExp_E} \times E$, is the \emph{guarded timed response relation},
    \item $\noresponserel,\milestonerel,\includerel,\excluderel,\valuerel\subseteq E \times \mathsf{BExp_E} \times E$ are the \emph{guarded cancel, milestone, include, exclude and value relations,} respectively,
    \item $L=\mathcal{P}(R) \times A$ is the \emph{set of labels}, with $R$ and $A$ sets of roles and actions,
    \item $l\colon E \to L$ is a \emph{labelling function} between events and labels.
\end{enumerate}
\end{definition}

The sub-process function $\spg(e)$ defines a partial containment relation of events, which allows an event to be refined by a sub-process defined by the events contained in it. We call such a refined event a sub-process event. A sub-process event gets executed when an event contained in it is executed, and no events of the sub-process in the resulting marking are required to be executed in the future.

   As already informally described above, the marking $M = (\Ex,\Re,\In,\Va)$ defines the state of the process. Formally, the marking consists of three partial functions ($\Ex$, $\Re$, and $\Va$) and a set $\In$ of events. $\Ex(e)$, if defined, yields the time since event $e$ was last \emph{{\bf Ex}ecuted}. $\Re(e)$, if defined, yields the deadline for when the event is \emph{{\bf Re}quired} to happen (if it is included). The set $\In$ is the currently {{\bf In}cluded events}. Finally, $\Va(e)$, if defined, is the current value of an event. 
  
\paragraph{Enabledness.} The condition $\conditionrel$ and milestone $\milestonerel$ relations constrain the enabling of events and determine when events can be executed. As exemplified above, a condition $e' \conditionrel e$ means that $e'$ must have been executed at least once or currently be excluded for $e$ to be enabled. A milestone $e' \milestonerel e$ means that $e'$ must either be currently excluded or not be pending for $e$ to be enabled. In the example in Fig.~\ref{fig:commit-and-reveal}, the milestone relations ensure that the \emph{commit} action cannot be repeated as long as required executions of \emph{reveal}, \emph{decide}, \emph{fail}, or \emph{pass} are pending.
Formally,
an event $e$ is enabled in marking $M=(\Ex,\Re,\In,\Va)$ and can be executed by role $r\in R$, if $l(e)=(R',a)$ for $r\in R'$ and 
(1) $e$ is included: $e\in \In$, 
(2) all conditions for the event are met: 
$\forall e'\in E. (e',k,d,e)\in \conditionrel. (e'\in \In \wedge \evalexp{d}{M})\implies \Ex(e')\geq k$ and 
(3) all milestones for the event are met: 
$\forall e'\in E. (e',k,d,e)\in \milestonerel. (e'\in \In \wedge \evalexp{d}{M})\implies \Re(e') \text{ is undefined}$ and
(4) $e$ is not contained in a sub-process event, or $\spg(e)$ is enabled and can be executed by role $r$.

In the DCR graph in Fig.~\ref{fig:commit-and-reveal}, the only enabled event is the event \emph{commit}. 
It is enabled because it is included and the source events of the two milestone rules are not initially required to be executed. The \emph{reveal} and \emph{decide} events are blocked by condition rules, and the \emph{fail} and \emph{pass} events are disabled because they are initially excluded (marked by a dashed border).

We refer the reader to~
\cite{dcrdata,dcrsub} for a more detailed definition and explanation of the execution semantics of timed DCR graphs with data and sub-processes.

\section{Smart Contract Design Patterns as DCR Graphs}\label{sec:dcr-patterns}

Due to the high stakes involved in applications, ensuring the safety and security of smart contracts is crucial. To address this, both the Solidity documentation and the developer and smart contract security community have put forth a range of recommendations. A considerable number of these recommendations are now known as \emph{design patterns}~\cite{gamma1995design}, because they are widely adopted as a solution to recurring design problems. These patterns promote the creation of contracts that are designed with safety and security in mind, mitigating potential risks and safeguarding users' assets in the design phase. 

We collected these design patterns from academic design pattern surveys  \cite{EmpiricalAnalysisSmart2017,governanceLiteratureReview,DesignPatterns2018,securityPatterns,SolidityPatternsWebsite}, documentation of Solidity \cite{solidity-web} and the Ethereum Foundation \cite{EthereumDevelopmentDocumentation}, and recommendations by a popular contract auditing company \cite{PrecautionsEthereum}. These design patterns are also confirmed by their occurrence in popular libraries and contracts such as OpenZeppelin, SolidState Solidity, and Aragon OSx \cite{openzeppelinLib,AragonOSxProtocol2023,SolidStateSolidity2023,CompoundProtocol2023,SmartcontractkitChainlink2023}.

\begin{figure}[t]
    \centering
    \includegraphics[scale=.5]{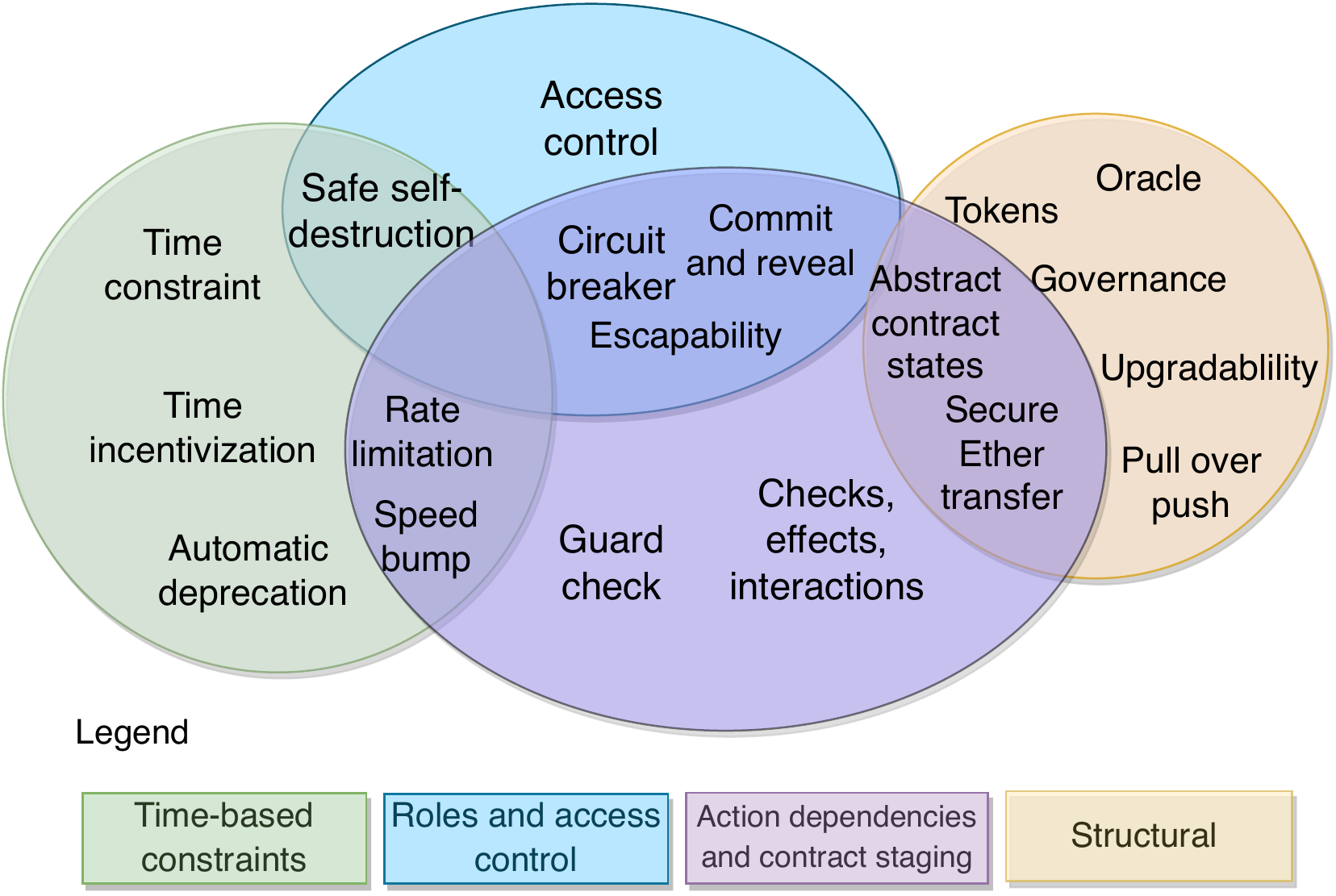}
    \caption{Classification of smart contract high-level design patterns (upper part of Table~\ref{table:patterntable})}
    \label{fig:states-ordering}
\end{figure}

\makeatletter
\newcommand{\thickhline}{%
    \noalign {\ifnum 0=`}\fi \hrule height 1pt
    \futurelet \reserved@a \@xhline
}
\newcolumntype{"}{@{\hskip\tabcolsep\vrule width 1pt\hskip\tabcolsep}}
\makeatother

\begin{table}[t]
\caption{Smart contract design patterns and their respective DCR graph models. High-level patterns (upper part of the table) are further categorized in Fig.~\ref{fig:states-ordering}.}
\label{table:patterntable}
\centering


\begin{threeparttable}
{\scriptsize

\begin{tabular}{lcl}
\textbf{Design Pattern} & \textbf{Libraries} & \multicolumn{1}{c}{\textbf{DCR Model}} \\
\thickhline
    \rowcolor{mygrey}
    \multicolumn{3}{c}{\textbf{High-level Patterns}}\\
    \hline

Time constraint
\cite{EmpiricalAnalysisSmart2017} & \cite{openzeppelinLib} & \href{https://github.com/mojtaba-eshghie/SmartContractDesignPatternsInDCRGraphs/blob/main/design-pattern-models/time-constraint.md}{GitHub}, \S\ref{time-constraint-dcr}\\

\hline

Time incentivization\tnote{1} & \cite{Augur2023, MakerProtocolWhite, CompoundProtocol2023, Aragoncourt2023, Synthetix} & \href{https://github.com/mojtaba-eshghie/SmartContractDesignPatternsInDCRGraphs/blob/main/design-pattern-models/time-incentivizing.md}{GitHub}, \S\ref{time-incent-dcr}, \S\ref{sec:casino}
\\

\hline

Automatic deprecation\cite{DesignPatterns2018} & --- & \href{https://github.com/mojtaba-eshghie/SmartContractDesignPatternsInDCRGraphs/blob/main/design-pattern-models/automatic-deprecation.md}{GitHub}, \S\ref{automatic-deprecation-dcr} \\
\hline

Rate limitation\cite{securityPatterns} & \cite{PrecautionsEthereum} & \href{https://github.com/mojtaba-eshghie/SmartContractDesignPatternsInDCRGraphs/blob/main/design-pattern-models/rate-limiting.md}{GitHub}, \S\ref{rate-limiting-dcr} \\

\hline

Speed bump\cite{securityPatterns} & \cite{PrecautionsEthereum} & \href{https://github.com/mojtaba-eshghie/SmartContractDesignPatternsInDCRGraphs/blob/main/design-pattern-models/speed-bump.md}{GitHub}, \S\ref{speedbump-dcr}\\

\hline

Safe self-destruction\cite{DesignPatterns2018, EmpiricalAnalysisSmart2017} & \cite{openzeppelinLib} & \href{https://github.com/mojtaba-eshghie/SmartContractDesignPatternsInDCRGraphs/blob/main/design-pattern-models/safe-self-destruction.md}{GitHub}, \S\ref{safe-self-destrcution-dcr}\\

\hline

Ownership / Authorization / Access Control  
\cite{EmpiricalAnalysisSmart2017,DesignPatterns2018,SolidityPatternsWebsite} & \cite{openzeppelinLib,SolidStateSolidity2023,AragonOSxProtocol2023}   & 
\href{https://github.com/mojtaba-eshghie/SmartContractDesignPatternsInDCRGraphs/blob/main/design-pattern-models/access-control.md}{GitHub}, \S\ref{access-control-dcr}, \S\ref{sec:casino}\\

\hline

Commit and reveal\cite{DesignPatterns2018} & --- & \href{https://github.com/mojtaba-eshghie/SmartContractDesignPatternsInDCRGraphs/blob/main/design-pattern-models/commit-and-reveal.md}{GitHub}, \S\ref{commit-and-reveal-dcr}, \S\ref{sec:casino}\\

\hline

Circuit breaker / Emergency stop\cite{securityPatterns} & --- & \href{https://github.com/mojtaba-eshghie/SmartContractDesignPatternsInDCRGraphs/blob/main/design-pattern-models/circuit-breaker.md}{GitHub}, \S\ref{circuit-breaker-dcr} \\
\hline

Escapability\tnote{1} & \cite{escapable, DecentralizedEscapeHatch, ImplementEscapeHatch, ExplainedAkutarsNFT} & \href{https://github.com/mojtaba-eshghie/SmartContractDesignPatternsInDCRGraphs/blob/main/design-pattern-models/escapability.md}{GitHub}, \S\ref{escapability-dcr} \\
\hline

Checks, effects, interactions\cite{securityPatterns,SolidityPatternsWebsite} & --- & \href{https://github.com/mojtaba-eshghie/SmartContractDesignPatternsInDCRGraphs/blob/main/design-pattern-models/checks-effects-interactions.md}{GitHub}, \S\ref{checks-effects-interactions-dcr}\\
\hline

Guard check\cite{SolidityPatternsWebsite} & \cite{hodlit} & \href{https://github.com/mojtaba-eshghie/SmartContractDesignPatternsInDCRGraphs/blob/main/design-pattern-models/guard-check.md}{GitHub}, \S\ref{guard-check-dcr} \\

\hline

Abstract contract states\cite{DesignPatterns2018} & \cite{openzeppelinLib} & \href{https://github.com/mojtaba-eshghie/SmartContractDesignPatternsInDCRGraphs/blob/main/design-pattern-models/abstract-contract-states.md}{GitHub}, \S\ref{fsm-dcr}, \S\ref{sec:casino}\\

\hline

Secure Ether transfer\cite{SolidityPatternsWebsite} & --- &  \href{https://github.com/mojtaba-eshghie/SmartContractDesignPatternsInDCRGraphs/blob/main/design-pattern-models/secure-ether-transfer.md}{GitHub}, \S\ref{secure-ether-transfer-dcr} \\
\hline

Oracle \cite{EmpiricalAnalysisSmart2017,DesignPatterns2018} & ---   & 
\href{https://github.com/mojtaba-eshghie/SmartContractDesignPatternsInDCRGraphs/blob/main/design-pattern-models/oracle.md}{GitHub}, \S\ref{oracle-dcr}\\
\hline

Token \cite{EmpiricalAnalysisSmart2017} & \cite{openzeppelinLib}   & 
\href{https://github.com/mojtaba-eshghie/SmartContractDesignPatternsInDCRGraphs/blob/main/design-pattern-models/tokens.md}{GitHub}, \S\ref{tokens-pattern-dcr} \\
\hline

Pull over push\cite{DesignPatterns2018} & \cite{openzeppelinLib} & \href{https://github.com/mojtaba-eshghie/SmartContractDesignPatternsInDCRGraphs/blob/main/design-pattern-models/pull-over-push.md}{GitHub}, \S\ref{pull-over-push-dcr}\\

\hline

Upgradability\cite{designPatternsGas,DesignPatterns2018} & \cite{openzeppelinLib} & \href{https://github.com/mojtaba-eshghie/SmartContractDesignPatternsInDCRGraphs/blob/main/design-pattern-models/upgradability.md}{GitHub}, \S\ref{upgradability-dcr}\\ 
\hline

Governance\cite{governanceLiteratureReview,EmpiricalAnalysisSmart2017} & \cite{Chainbridgesolidity2022a, CompoundV2Docs, openzeppelinLib} & \href{https://github.com/mojtaba-eshghie/SmartContractDesignPatternsInDCRGraphs/blob/main/design-pattern-models/governance.md}{GitHub}, \S\ref{governance-dcr} \\

\thickhline

\rowcolor{mygrey}
    \multicolumn{3}{c}{\textbf{Low-level Patterns}}\\
    \hline

Randomness \cite{EmpiricalAnalysisSmart2017,SolidityPatternsWebsite} & --- & \multicolumn{1}{c}{\xmark}\\

\hline

Safe math operations\cite{EmpiricalAnalysisSmart2017} & \cite{openzeppelinLib} & \multicolumn{1}{c}{\xmark}\\

\hline

Variable Packing\cite{designPatternsGas,SolidityPatternsWebsite} & \cite{openzeppelinLib} & \multicolumn{1}{c}{\xmark} \\
\hline

Avoiding on-chain data storage\cite{designPatternsGas} & \cite{openzeppelinLib} & \multicolumn{1}{c}{\xmark} \\

\hline

Mutex\cite{securityPatterns} & \cite{openzeppelinLib} &  \multicolumn{1}{c}{\xmark} \\ 

\hline

Freeing storage \cite{designPatternsGas} & --- & \multicolumn{1}{c}{\xmark} \\

\end{tabular}


\begin{tablenotes}
\item[1] We identify these as design patterns since they have been used as a recurring solution in several real-world smart contracts but have not yet been considered a design pattern in the literature. 
\end{tablenotes}

}
\end{threeparttable}
\end{table}

First, we identify the design patterns representing high-level behavior rather than implementation- and platform-specific patterns (Table \ref{table:patterntable}). The latter concerns features inside a function (the execution of which we model as an event in DCR graphs). The analysis of low-level patterns is orthogonal to our work and can be handled, e.\,g., by runtime analysis of code~\cite{EASYFLOW}. We then classify the high-level patterns into the following four categories (see Fig.~\ref{fig:states-ordering}):
\begin{enumerate}
    \item \textbf{Time-based constraints}: Time-based patterns impose constraints on when activities can be performed, which typically include deadlines and delays. 
    \item \textbf{Roles and access control}: Role-based access control~\cite{sandhu1998role} restricts access to given functions to predefined roles.
    \item \textbf{Action dependencies and contract staging}: High-level design of a smart contract may impose an ordering on any pair of activities.
    \item \textbf{Structural patterns}: These patterns impose a certain structure on the contract business process (and the implementation as a result) and are created by combining other design patterns. 
\end{enumerate}
Many patterns combine aspects of several categories; Fig.~\ref{fig:states-ordering} depicts a classification of 19 design patterns we have identified. We elucidate these patterns further below.
Also, we describe DCR graphs for selected design patterns here; the others are available on GitHub.\footnote{\url{https://github.com/mojtaba-eshghie/SmartContractDesignPatternsInDCRGraphs}}
Table \ref{table:patterntable} gives an overview of references of the design patterns, libraries that implement them, and their respective DCR models. 

In the following subsections, we delve into each design pattern, highlighting its utility, and, for a chosen subset, offer the visual representation of their model and a succinct description of the associated DCR graph models. This study provides supplementary details and examples for each pattern, along with the DCR model semantics used, in our corresponding GitHub repository. We plan to focus on comprehensive guidelines for smart contract modeling in future research.


\begin{myenv}{}

\myitem{\bf Time constraint.}\label{time-constraint-dcr}
In multi-stage business processes, code execution must adhere to specific stages. This can be achieved through time-based or action-based dependencies. The former denotes stages solely based on elapsed time~\cite{EmpiricalAnalysisSmart2017}. 
This pattern prohibits calling a function until a specific time is reached on the blockchain, represented by a delayed condition relation in DCR graphs. The simplest form of this pattern is modeled directly using a delayed condition DCR relation. Modeling more complex time constraints where only part of a function should be executed or blocked based on time is challenging and may require multiple guard conditions in DCR graphs. One approach is to interpret Solidity's \texttt{require} statements as guard conditions in DCR graphs, connecting multiple activities to shape the business logic. 

\myitem{\bf Time Incentivization.}\label{time-incent-dcr}
In Ethereum, smart contracts work as a reactive system where specific function calls execute transactions.
There are scenarios where certain actions should be performed at a specific time or when a specific condition is fulfilled.
A lack of action may prevent progress, opening up adversarial behavior (e.\,g., the eternal locking of assets).
The purpose of the \emph{time incentivization} pattern
is to motivate parties to cooperate even in the existence of conflicting interests. The incentivization is typically done by stipulating a deadline before which an actor shall make a move. The actor that misses the deadline can afterward be punished by other actors, e.\,g., by forfeiting the bets, as modeled in the casino contract (see \S\ref{sec:casino}).

To demonstrate this pattern, we use the simple example of giving a loan and then motivating the client to pay for the loan. Giving a loan is performed only by \emph{bank} role. In Fig.~\ref{fig:time-incent-dcr}, immediately after giving the loan, the bank includes $\includerel$ both the \emph{pay loan} and \emph{fine} activities. Without any more relationships, this would mean that the bank might increase the interest on the loan without giving the client enough time to pay for it. This issue is resolved by using the pre-condition arrow $\conditionrel$ from \emph{give loan} to activity \emph{fine}. As this pre-condition arrow has a deadline attribute of P1M (one month), it will suspend the availability of \emph{fine} to one month later. Without this pattern, the client could refuse to pay the loan by not participating in any further transaction. 

Despite the widespread usage of this pattern in popular smart contracts such as Augur, MakerDAO, Compound, Aragon Court, and Synthetix~\cite{Augur2023, MakerProtocolWhite, CompoundProtocol2023, Aragoncourt2023, Synthetix} to incentivize taking the next step by the actor(s), the current work is the first one classifying it as a design pattern and formalizing it using DCR graphs.


\begin{figure}[t]
    \centering
    \includegraphics[width=0.5\textwidth]{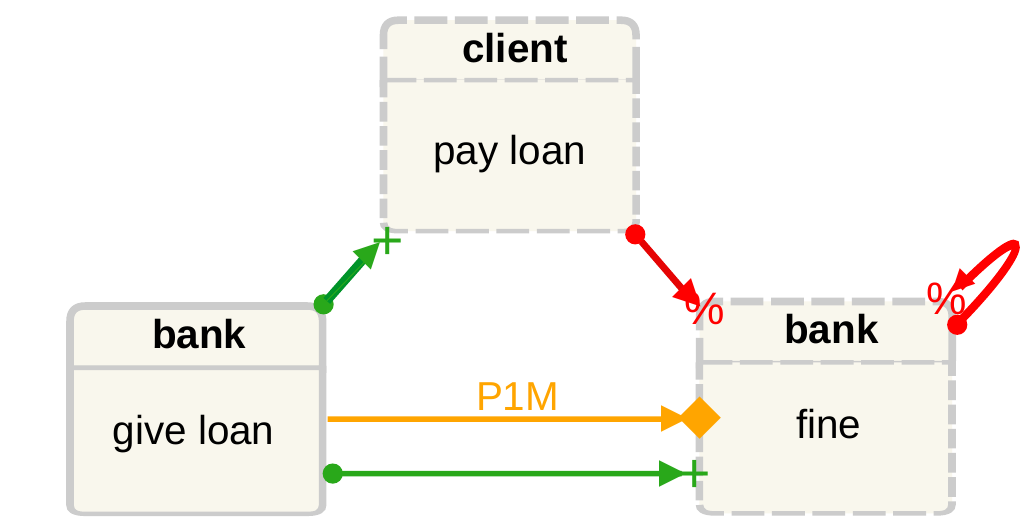}
    \caption{DCR model of time incentivization design pattern.}
    \label{fig:time-incent-dcr}
\end{figure}

\myitem{\bf Automatic Deprecation.}\label{automatic-deprecation-dcr}
Automatic deprecation is the opposite of a time constraint, 
stipulating a deprecation time (block number) after which a function is not executable anymore \cite{DesignPatterns2018}. 
In Solidity code, such functions are typically enabled by a \emph{require} statement checking at the function entry point against the expiration. This means that a smart contract function can be called and reverted, which is different from DCR model semantics, where an activity is enabled only if it will successfully execute. We model this in DCR by checking the deprecation condition on an exclusion arrow from another activity to the target activity subject to deprecation.

\myitem{\bf Rate Limitation.}\label{rate-limiting-dcr}
Rate limitation imposes a limit on the number of successful function calls by a participating user during a specific time period  \cite{PrecautionsEthereum}. The more common type of this pattern that we analyze here explicitly limits the total amount of transfers allowed during the defined period. 
%



To model this pattern, we assume the sensitive activity is the \emph{withdraw} operation. We present the model in Fig. \ref{fig:rate-limiting}. When the model is simulated, the only included available activity to execute is \emph{set limit}. The \emph{new period} activity is initially \emph{executed} (tick the on activity box). The gray arrow $valuerel$ from \emph{new period} to \emph{rate limiter} copies value $0$ to \emph{rate limiter} every time \emph{new period} is executed.
Each execution of \emph{new period} sets a deadline and delay of one day (P1D) for the given activity. Having such relationships (response and precondition) on \emph{new period} and assigning an automatic agent to the \emph{system} (when simulating the model) ensures that \emph{new period} is indeed executed at exactly one-day intervals. In Fig.~\ref{fig:rate-limiting}, labels P1D on the
reflexive pre-condition $\conditionrel$ and
reflexive response $\responserel$
arrows of \emph{new period} impose this periodic execution. Based on the purple milestone relation $\milestonerel$, if the current period amount does not exceed the limit, role \emph{user} can withdraw. Furthermore, having the milestone relation from \emph{new period} to \emph{rate limiter}  occur periodically with $\mathit{currentamount} \ge \mathit{limit}$ ensures that if the current withdrawal of the period exceeds the limit, \emph{withdraw} will not be executable until the next execution of \emph{new period}. The new execution of \emph{new period} resets the \emph{currentamount} to $0$ again.


\begin{figure}[t]
    \centering
    \includegraphics[scale=0.67]{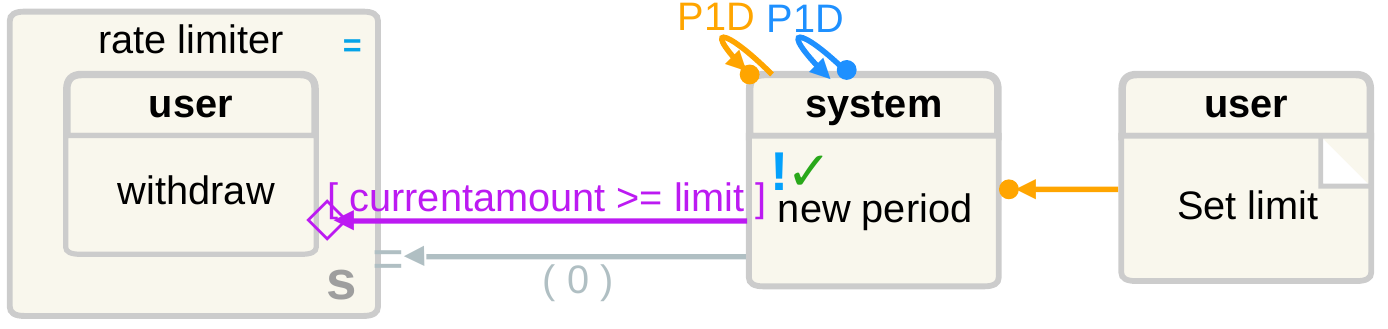}
    \caption{Rate limitation pattern modeled in DCR graphs}
    \label{fig:rate-limiting}
\end{figure}
 
\myitem{\bf Timed Temporal Constraint (Speed Bump).}\label{speedbump-dcr}
A \emph{speed bump} is used to slow down critical operations such as the withdrawal of assets, authorization of significant actions, etc. \cite{PrecautionsEthereum}. It imposes a temporal barrier that gives enough time to a monitoring system to detect a problematic activity and mitigate it. This pattern is a specialized form of the time constraint pattern where the participating user can only execute an action
after a predefined time period has passed (from the point the action request has been registered).
%
The wait time is modeled using a \emph{delay} on a condition arrow from the activity requesting the specified action to the actual action.


\myitem{\bf Safe Self-Destruction.}\label{safe-self-destrcution-dcr}
It is possible to define a function in Solidity that uses \lstinline{selfdestruct(address target)} to destroy the contract intentionally and send all Ether in the contract account to the target. Safe self-destruction 
is about limiting the execution of the function to
specific roles such as the administrator. \cite{DesignPatterns2018, EmpiricalAnalysisSmart2017}.
The simplest way to achieve this is to refine the access control pattern (\S \ref{access-control-dcr}). 
However, guard check and time constraint patterns (\S \ref{guard-check-dcr} and \S \ref{time-constraint-dcr}) can also be used to ensure safety. 

\myitem{\bf Access Control.}\label{access-control-dcr}
Access control restricts access to desired functions to only a subset of accounts calling them \cite{EmpiricalAnalysisSmart2017,DesignPatterns2018,SolidityPatternsWebsite}. A common instance of this pattern is to initialize a variable \emph{owner} to the contract deployer and only allow this account to successfully call certain functions.
Here, we can nicely exploit that access control is built into DCR graphs as a first-class citizen, in the form of roles assigned to activities. Each activity in a DCR model can be limited to one or more specific roles.
In simpler scenarios, roles are assigned statically to accounts when a contract is deployed on the blockchain.
In general, however, access rights can be assigned dynamically.
DCR graphs support this using \emph{activity effects} from an external database source. This feature allows changing the roles of activities as a result of an activity being executed.

\myitem{\bf Commit and Reveal.}\label{commit-and-reveal-dcr}
In a public permissionless blockchain platform such as Ethereum, transaction data is public \cite{DesignPatterns2018}. Therefore, if a secret is sent along with a transaction request, participants in the blockchain consensus protocol can see the secret value even before the transaction is finalized. On the other hand, the party holding the secret should commit to it
before other parties act, so the secret cannot be changed after the fact.
The commit and reveal pattern addresses this problem and works in two phases. In Phase 1, a piece of data is submitted that depends on the secret (which itself is not yet submitted). Often, that data is the crypto-hash of the secret, such that the secret cannot be reconstructed. Phase 2 is the submission (and reveal) of the secret itself. 
%
We use a combination of condition, milestone, and response relations to enforce the ordering of actions in the commit and reveal pattern in Fig.~\ref{fig:commit-and-reveal}.
Here, the activity \emph{reveal} is blocked
initially by the condition relation from \emph{commit} to \emph{reveal}
, and is enabled 
once a user commits. The commit makes \emph{reveal} pending (by the response relation arrow). 
Finally, the milestone relation from the pending \emph{reveal} to \emph{commit} means that unless a reveal happens, no other commit is possible. The decision is then made using the \emph{decide} activity based on committed and revealed values. 

\myitem{\bf Circuit Breaker (Emergency Stop).}\label{circuit-breaker-dcr}
This pattern enables the contract owner to temporarily halt the contract's normal operations until a manual or automatic investigation is performed \cite{securityPatterns}. Other contract functions, such as those based on timed temporal constraints (\S\ref{speedbump-dcr}), can also trigger the circuit breaker.
%
To model this design pattern, we categorize activities into two subsets: activities that are available in the normal execution of the contract and those that are only available when the circuit breaker is triggered. There is a milestone relationship $\milestonerel$ between circuit breaker grouping and all other DCR nodes. The existence of this milestone helps to disable the execution of all of these activities by making the circuit breaker pending. In Fig.~\ref{fig:circuit-breaker-dcr}, the activity \emph{panic} executed by the \emph{monitor} role makes the circuit breaker pending (\emph{panic} $\responserel$ \emph{circuit breaker}). This means unless \emph{revive} activity in the circuit breaker group is executed, none of the \emph{buy}, \emph{sell}, \emph{transfer}, and \emph{panic} activities are executable. 
Executing \emph{contingency} instead will enable a contingency plan (related to \S \ref{escapability-dcr}).

\begin{figure}[t]
    \centering
    \includegraphics[scale=0.47]{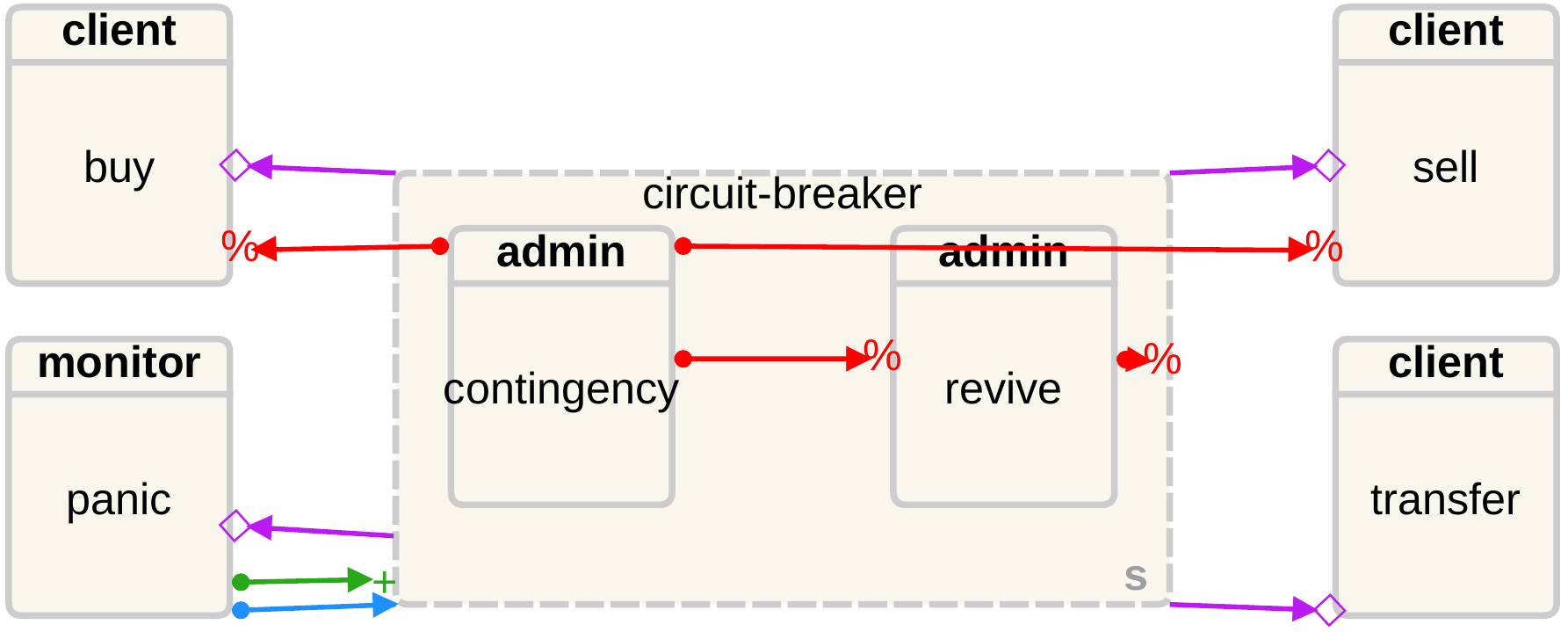}
    \caption{Circuit breaker design pattern DCR model}
    \label{fig:circuit-breaker-dcr}
\end{figure}

\myitem{\bf Escapability.}\label{escapability-dcr}
There have been cases where a vulnerability in the contract triggered by a certain transaction led to funds being locked in the contract~\cite{ExplainedAkutarsNFT}. 
%
To prevent this, a smart contract can have a function whose logic is independent of the main contract logic; when triggered, it can withdraw all assets in the contract to a certain address. This new address can be the upgraded version of the contract that contains a patch for the vulnerability. 
Escapability is arguably the complementary pattern for the circuit breaker pattern (\S \ref{circuit-breaker-dcr}), as it concerns the functionality behind the \emph{contingency} activity in Fig.~\ref{fig:circuit-breaker-dcr} for the circuit breaker. This functionality often consists of transferring assets to an escape hatch. Despite being used by the community \cite{DecentralizedEscapeHatch, ImplementEscapeHatch, ExplainedAkutarsNFT}, the current work is the first one promoting it as a design pattern. 

\myitem{\bf Checks, Effects, Interactions.}\label{checks-effects-interactions-dcr}
This pattern is concerned with the order of certain activities, especially when interactions with other contracts (external calls) happen \cite{securityPatterns,SolidityPatternsWebsite}. External calls can be risky, as call targets cannot necessarily be trusted. One risk is that the called contract calls back into the calling contract before returning, purposefully abusing the calling contract's logic \cite{dynamicVul}. To prevent such exploits, 
the caller first performs \emph{checks} on its bookkeeping variables (variables keeping the balance of tokens, assets, etc.). Then, it modifies these bookkeeping variables based on the business logic \emph{(effects)}. Last, there are \emph{interactions} with (i.\,e., calls to) other contracts.
In DCR graphs, we specify this strict ordering via inclusion/exclusion relations among the respective activities.

%

\myitem{\bf Guard Check.}\label{guard-check-dcr}
A guard check validates user inputs and checks bookkeeping variables and invariants before the execution of the function body (mainly as a \emph{require} statement) \cite{hodlit}. This pattern is often applied using function {modifiers} in Solidity and
represented using guard conditions on DCR relations. 

\myitem{\bf Abstract Contract States.}\label{fsm-dcr}
In most processes,
action dependencies impose a partial order on action executions
that a smart contract has to follow, as shown in the casino contract (\S \ref{sec:casino}). In Solidity, a state variable of type enumeration can mimic a finite state automaton \cite{DesignPatterns2018}, whose state transitions enforce the set of executable functions, encoding a partial order among action executions.

In DCR graphs, such dependencies (partial orderings of actions) can be represented explicitly.
If the ordering between activities does not matter, no arrows are required. Therefore, DCR graphs can make contract states obsolete \emph{at the modeling level.}
If there is a strong reason for modeling the abstract contract states instead of the action dependencies they imply, it is still possible to model them using DCR graphs. This is done by grouping activities of the same state into the same group in DCR graphs and using arrows between state groupings to reflect state transitions of the system.

\myitem{\bf Secure Ether Transfer.}\label{secure-ether-transfer-dcr}
This structural design pattern imposes a design choice between various ways of sending Ether from one contract to the other (via \emph{send}, \emph{transfer}, or \emph{call}) \cite{SolidityPatternsWebsite}. Using each of them requires a distinct way of ensuring the target contract cannot harm the contract sending Ether.
%
%
As a structural design pattern, Secure Ether Transfer imposes certain guard checks, mutual exclusions, and ordering of actions to ensure that an external call (especially to transfer Ether) is not exploitable by a malicious party. Therefore, this pattern can be represented in DCR graphs as action dependency relation in combination with the guard check (\S\ref{guard-check-dcr}) and mutex (Table \ref{table:patterntable}) design patterns. 

\myitem{\bf Oracle.}\label{oracle-dcr}
Oracles enable smart contracts to incorporate off-chain data in their execution and push information from a blockchain to external systems \cite{EmpiricalAnalysisSmart2017,DesignPatterns2018}. 
%
The oracle pattern employs an external call to another \emph{service smart contract} (data source) to register the request for off-chain data. This registration call information should also be kept in bookkeeping variables inside the contract itself. When the data is ready in the service contract, it will inform the main contract about the result by calling a specific \emph{callback} function.
%
To model this, the callback function of the smart contract is excluded by default and is included when the smart contract calls an oracle. 

\myitem{\bf Token Design Patterns.}\label{tokens-pattern-dcr}
Tokens represent assets, their behavior, and manageability \cite{EmpiricalAnalysisSmart2017}. Ethereum smart contracts and token standards (such as ERC-20, ERC-721, and ERC-777) enable developers to use tokens according to specific requirements.
DCR graphs can model both tokens 
and their interacting contracts. The ERC-20 token standard model included in the accompanying repository to this work involves inclusion/exclusion relations to model the partial ordering of activities. Tokens and contracts that use this model typically involve several other design patterns (most notably \S \ref{pull-over-push-dcr} and \S \ref{secure-ether-transfer-dcr}).  

\myitem{\bf Pull Over Push.}\label{pull-over-push-dcr}
A contract might need to send a token or Ether to other accounts. The ``pull over push'' pattern discourages pushing tokens or Ether to the destination as a side-effect of calling a function. Rather, it encourages exposing a \emph{withdraw} function that users of the contract can call \cite{DesignPatterns2018} for this reason. This inclination towards pull is based on the fact that when sending Ether or tokens via any external call (even when adhering to patterns such as \S \ref{secure-ether-transfer-dcr}), the receiver may act unexpectedly before returning control.
We model this pattern in a DCR graph by having an extra activity for the \emph{withdraw} functionality. 

\myitem{\bf Upgradability.}\label{upgradability-dcr}
This design pattern consists of up to five parts:
(1)~The proxy keeps addresses of referred contracts. 
(2)~The data segregation part separates the logic and data layers by storing data in a separate smart contract.
(3)~The satellite part outsources functional units to separate satellite contracts and stores their addresses in a base contract, allowing the replacement of their functionality. 
(4)~The register contract tracks different versions of a contract and points to the latest one.
(5)~While keeping the old contract address, the relay pattern uses a proxy contract to forward calls and data to the newest contract version~\cite{DesignPatterns2018}. 
Data segregation, satellite, and relay are platform-dependent low-level patterns, which we do not capture with our DCR graph model. Our upgradability pattern model (Table \ref{table:patterntable}) instead explicitly includes activities for the register and proxy parts.

\myitem{\bf Governance.}\label{governance-dcr}
On-chain governance is a crucial component of decentralized protocols, allowing for decision-making on parameter changes, upgrades, and management \cite{governanceLiteratureReview, EmpiricalAnalysisSmart2017}. 
The governance pattern is typically used to allow token holders or a group of privileged users to vote on proposals and make decisions that affect the contract's behavior. 
%
This pattern works in conjunction with other patterns, such as guard check (\S \ref{guard-check-dcr}) and role-based access control (\S \ref{access-control-dcr}).

\end{myenv}

\section{Modeling and Analysis of A Casino Smart Contract}\label{sec:casino}
As an example of how patterns modeled in DCR graphs come into play when modeling a concrete smart contract scenario, we present a simple casino contract~\cite{casinosource}.\footnote{The scenario was originally provided by Gordon Pace.} It uses four design patterns identified in Table~\ref{table:patterntable}: time incentivization, role-based access control, commit and reveal, and abstract contract states. This endeavor demonstrates how utilizing and combining the DCR model of several design patterns into one model captures the intended smart contract design.

The casino has two explicitly declared roles, \textit{operator} and \textit{player}. It also contains three abstract states (see \S\ref{fsm-dcr}): \textit{IDLE}, \textit{GAME\_AVAILABLE}, and \textit{BET\_PLACED}. Three modifiers check the pre-conditions $\conditionrel$ of each function based on the roles and the state the contract is in.

\begin{figure}[t]
    \centering
    \includegraphics[scale=0.47]{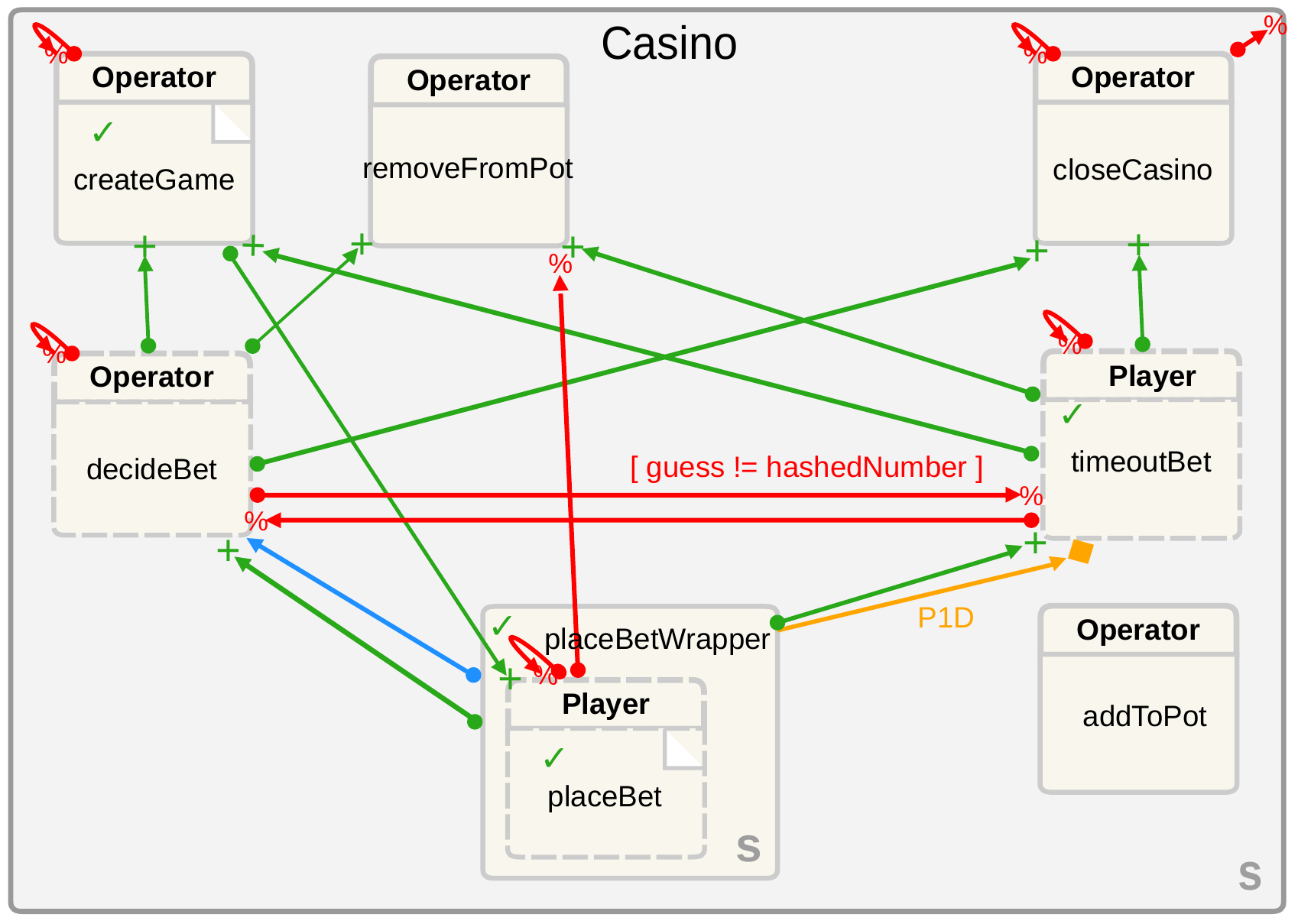}
    \caption{Casino contract model.}
    \label{fig:casino-final}
\end{figure}

Fig.~\ref{fig:casino-final} shows the DCR model of this contract. The activities all reflect functions of the same name, except subprocess \textit{casino}, which everything is grouped under. This subprocess reflects the behavior of the deployed contract, which includes a suicidal \textit{closeCasino} function that \textit{selfdestructs},
shown by an exclusion arrow from \textit{closeCasino} to the subprocess in Fig.~\ref{fig:casino-final}.
Without a subprocess, an exclusion arrow would go from \textit{closeCasino} to all other activities, which is visually unappealing. Furthermore, we do not model the actual states of the contract, instead choosing to order the activities by inclusion $\includerel$ and exclusion $\excluderel$ arrows.

When the casino contract is deployed, it is in the \textit{IDLE} abstract state. It is possible to create a game, add to the pot, remove from it, or self-destruct. Creating a game will change the abstract state to \textit{GAME\_AVAILABLE}, which enables anyone in the Ethereum network to place a bet and take the role of the player (as a result). The function \textit{decideBet} checks if the player is the winner by comparing the guess with the secret number. This gives both the player and the operator a $50\,\%$ chance of winning the game. In the model, a response arrow $\responserel$ from \emph{placeBet} to \emph{decideBet} emphasizes that \emph{decideBet} has to execute at some point and should not block the game from continuing. However, since continuing the game at this point depends on the operator making a transaction, it is possible for a malicious operator or buggy reverted \emph{decideBet} function to lock the funds the player puts in the game. Furthermore, after a player places the bet, the operator should not be able to change the actual secret stored. Therefore, three patterns are used in the model to provide the following functionality:

\begin{itemize}
    \item A time incentivization pattern (\S\ref{time-incent-dcr}) ensures that continuing the game is the favorable option for the casino operator.
Fig.~\ref{fig:casino-final} shows the required mechanism, where \textit{timeoutBet} becomes available with a desirable delay (here P1D, one day) to provide the player with an option when the operator is unable or unwilling to make a transaction to \textit{decideBet}. Calling this function after the timeout guarantees the player wins the game and motivates the operator to decide the game in time.
    \item A commit and reveal pattern (\S\ref{commit-and-reveal-dcr}) is used to ensure when operator \emph{createGame} is called, the operator commits to a secret without sending it. The revealing phase of this pattern is performed in \emph{decideBet}, where the secret is submitted, checked, and compared to the player's guess.
    \item A role-based access control pattern (\S\ref{access-control-dcr}) to confine \emph{player} and \emph{operator} roles to their respective activities.
\end{itemize}
The abstract contract states pattern (\S \ref{fsm-dcr}) used in the implementation (Fig.~\ref{fig:solcode1}) is not needed in the model (Fig. \ref{fig:casino-final}): The model's partial ordering provides the same semantics without the complications of abstract contract states.

As mentioned in \S \ref{sec:intro}, DCR specifications provide a basis for automated analysis, to \emph{verify} that the implementation of the contracts adheres to their models.

We have implemented a runtime monitoring tool ``Clawk''~\cite{clawk}. 
Clawk captures transactions from the Ethereum client and executes an instance of the DCR graph model in tandem. For each Ethereum transaction, Clawk checks if the DCR graph model permits a corresponding action in the model. If this is not the case, the tool reports a violation.

By leveraging runtime information, our framework enables automated runtime verification. While runtime verification in the blockchain domain is typically associated with the performance overhead of contract or platform instrumentation \cite{EVMStar,EASYFLOW,ECF}, we address this concern by placing the monitor off-chain. If any deviations from the specification are detected, Clawk generates alerts, which can be used to enhance the contract's implementation or enable a circuit breaker pattern in the contract implementation (cf.~\S \ref{circuit-breaker-dcr}). 

\section{Related Work}\label{sec:related-work}

Smart contracts often involve multiple dependent transactions, a challenge that has been addressed through various approaches. Sergey and Hobor discuss the non-deterministic nature of transaction ordering decided by miners~\cite{concurrentPersSC}, while other works focus on commutativity conditions to exploit interleavings~\cite{autoGenCommutativeCond} or identify serializable transactions in Ethereum~\cite{addingConcurrency2SC}. These issues have also been modeled using finite state automata, which can lead to ``bad states'' in certain scenarios (e.\,g., when most of the actions are not accessible)~\cite{rvethereum}. DCR graphs offer a more elegant solution in such cases. Transactions and their dependencies can be graphically represented, as demonstrated by Chen et al., who use this to identify potential security issues~\cite{chen2020understanding}. Our work uniquely combines these general properties~\cite{wang-2022} with specific features like access control~\cite{liu-2022} to provide a comprehensive framework for smart contracts.


Chen et al. use graphs to analyze transactional dependencies and security in smart contracts, but their approach is statistical and less precise than ours~\cite{chen2020understanding}. While general properties of smart contracts focus on transactional integrity (not creating or destroying funds in the contract)~\cite{wang-2022}, specific features can be modeled through access control~\cite{liu-2022} or finite-state machines~\cite{guth2018specification}. 
To our knowledge, our work is the first to systematically apply a combination of these two aspects to smart contracts in terms of general and reusable patterns.

\section{Conclusion and Future Work}\label{sec:conclusion}
Smart contracts are critical yet complex pieces of software that encode business processes in an executable form on a blockchain. We collected 19 smart contract design patterns that dissect complex smart contracts into smaller reusable components, making it easier to reason about them. DCR graphs are an ideal way to formally model the semantics of these patterns, supporting the concepts of time, data, and sub-processes. We demonstrate their usefulness on the casino smart contract that combines multiple patterns.

The contract DCR models serve as a repository of reusable templates for developing more secure and efficient smart contracts across various applications and smart contract execution platforms. This not only aids in the initial design phase but also has uses in monitoring the contract behavior, allowing for automated verification~\cite{clawk}, which reduces the risk of vulnerabilities. Future directions of our research include an extensive evaluation of the Clawk tool, combinations of static or dynamic analysis of low-level patterns~\cite{EASYFLOW} with runtime monitoring against our high-level design patterns, as well as automated discovery of the models from the contract transaction history.  


\newpage
\section*{Appendix}
\appendix

\renewcommand{\lstlistingname}{Appendix Fig.}
\renewcommand{\thelstlisting}{\thesection-\arabic{lstlisting}}

\section{Casino Contract Source}

\begin{lstlisting}[basicstyle=\fontsize{8}{10}\ttfamily,numberstyle=\tiny,language=Solidity,escapechar=|,caption=The complete source code for Casino contract]
pragma solidity ^0.4.11;

contract Casino {
    address public operator;
    uint256 public timeout;
    uint256 constant DEFAULT_TIMEOUT = 30 minutes;
    uint256 public pot;
    bytes32 public hashedNumber;
    address public player;

    enum Coin {
        HEADS,
        TAILS
    }
    struct Wager {
        uint256 bet;
        Coin guess;
        uint256 timestamp;
    }
    Wager private wager;
    enum State {
        IDLE,
        GAME_AVAILABLE,
        BET_PLACED
    }
    State private state;
    
    // Modifiers
    modifier inState(State _state) {
        require(_state == state);
        _;
    }
    modifier byOperator() {
        require(msg.sender == operator);
        _;
    }
    modifier noActiveBet() {
        require(state == State.IDLE || state == State.GAME_AVAILABLE);
        _;
    }
    // -----------------------------------------
    // Create a new casino
    constructor() public {
        operator = msg.sender;
        state = State.IDLE;
        timeout = DEFAULT_TIMEOUT;
        pot = 0;
        wager.bet = 0;
    }
    // Changing the timeout value
    function updateTimeout(uint256 _timeout) public byOperator noActiveBet {
        timeout = _timeout;
    }
    // Add money to pot
    function addToPot() public payable byOperator {
        // The operator can choose a positive value to pay and raise the pot by
        require(msg.value > 0);

        pot = pot + msg.value;
    }
    // Remove money from pot
    function removeFromPot(uint256 amount) public byOperator noActiveBet {
        require(amount > 0 && amount <= pot);
        pot = pot - amount;
        msg.sender.transfer(amount);
    }
    // Operator opens a bet
    function createGame(bytes32 _hashedNumber)
        public
        byOperator
        inState(State.IDLE)
    {
        hashedNumber = _hashedNumber;
        state = State.GAME_AVAILABLE;
    }
    // Player places a bet
    function placeBet(Coin _guess)
        public
        payable
        inState(State.GAME_AVAILABLE)
    {
        require(msg.sender != operator);
        require(msg.value > 0 && msg.value <= pot);
        state = State.BET_PLACED;
        player = msg.sender;
        wager = Wager({bet: msg.value, guess: _guess, timestamp: now});
    }
    // Operator resolves a bet
    function decideBet(uint256 secretNumber)
        public
        byOperator
        inState(State.BET_PLACED)
    {
        require(hashedNumber == keccak256(secretNumber));
        Coin secret = (secretNumber % 2 == 0) ? Coin.HEADS : Coin.TAILS;
        if (secret == wager.guess) {
            playerWins();
        } else {
            operatorWins();
        }
        state = State.IDLE;
    }
    // Player resolves a bet because of operator not acting on time
    function timeoutBet() public inState(State.BET_PLACED) {
        require(msg.sender == player);
        require(now - wager.timestamp > timeout);
        playerWins();
        state = State.IDLE;
    }
    // Player wins and gets back twice his original wager
    function playerWins() private {
        pot = pot - wager.bet;
        player.transfer(wager.bet * 2);
        wager.bet = 0;
    }
    // Operator wins, transferring the wager to the pot
    function operatorWins() private {
        pot = pot + wager.bet;
        wager.bet = 0;
    }
    // Operator closes casino
    function closeCasino() public inState(State.IDLE) byOperator {
        selfdestruct(operator);
    }
    function() {}
}
\end{lstlisting}

\section{Casino Contract Description}
The Casino contract simulates a simple coin-tossing game where an \emph{operator} and a \emph{player} interact. The contract starts in an \emph{IDLE} state, allowing the operator to create a game, add or remove funds from the pot, or even close the casino. Once a game is created, it transitions to the \emph{GAME\_AVAILABLE} state, enabling players to place bets. After a bet is placed, the contract moves to the \emph{BET\_PLACED} state, where the operator must decide the outcome of the bet. The contract then returns to the \emph{IDLE} state, ready for the next game.

\subsection{Design Patterns}
\subsubsection{Time Incentivization}
The contract employs a time incentivization pattern to ensure that the operator acts within a reasonable time frame. This is implemented through the timeout variable and the \emph{timeoutBet()} function. If the operator fails to decide the bet within the specified timeout, the player can call \emph{timeoutBet()} to automatically win the game. This mechanism encourages the operator to act promptly, mitigating the risk of funds being locked in the contract.

\subsubsection{Role-Based Access Control}
The contract uses a role-based access control pattern to restrict access to certain functions based on the caller's role. This is implemented using the \emph{byOperator} and \emph{inState} modifiers. The \emph{byOperator} modifier ensures that only the operator can call certain functions like \emph{createGame}, \emph{addToPot}, and \emph{removeFromPot}. The \emph{inState} modifier checks that the contract is in the appropriate state for the function to be called, effectively acting as a role-based control for the contract's states.

\subsubsection{Commit and Reveal}
The commit and reveal pattern is used to maintain the secrecy of the operator's choice until the bet is decided. The operator initially submits a hashed number using \emph{createGame}, which commits them to a secret number without revealing it. Later, in \emph{decideBet}, the operator reveals the secret number, which is then hashed and checked against the initially submitted hash. This ensures fairness and prevents the operator from changing their choice midway through the game.

\subsubsection{Abstract Contract States}
The contract uses an enumerated type, State, to represent its abstract states: \emph{IDLE}, \emph{GAME\_AVAILABLE}, and \emph{BET\_PLACED}. These states are managed by the private variable state and checked by the \emph{inState} modifier before executing certain functions. This pattern simplifies the contract's logic and makes it easier to understand the allowed transitions between states.

\subsection{Functions and Modifiers}
\subsubsection{Modifiers}
\begin{itemize}
    \item \emph{inState(State \_state)}: Checks if the contract is in the specified state.
    \item \emph{byOperator()}: Checks if the message sender is the operator.
    \item \emph{noActiveBet()}: Checks if there is no active bet.
\end{itemize}
    
\subsubsection{Core Functions}
\begin{itemize}
    \item \emph{constructor()}: Initializes the contract, setting the operator and default values.
    \item \emph{updateTimeout(uint256 \_timeout)}: Allows the operator to update the timeout value.
    \item \emph{addToPot()}: Allows the operator to add funds to the pot.
    \item \emph{removeFromPot(uint256 amount)}: Allows the operator to remove funds from the pot.
    \item \emph{createGame(bytes32 \_hashedNumber)}: Allows the operator to start a new game.
    \item \emph{placeBet(Coin \_guess)}: Allows a player to place a bet.
    \item \emph{decideBet(uint256 secretNumber)}: Allows the operator to decide the outcome of a bet.
    \item \emph{timeoutBet()}: Allows the player to win by default if the operator fails to act within the timeout.
    \item \emph{playerWins()}: Private function to handle the logic when the player wins.
    \item \emph{operatorWins()}: Private function to handle the logic when the operator wins.
    \item \emph{closeCasino()}: Allows the operator to close the casino and retrieve the remaining funds.
\end{itemize}

\bibliographystyle{splncs04}
\bibliography{refs}

\end{sloppypar}
\end{document}